\documentclass[12pt,preprint]{aastex}

\shorttitle{Photometric Variability of UM\,673}
\shortauthors{Sinachopoulos et al.}

\begin{document}

\title{Photometric Variability of the Gravitational Lens 0142-100 (UM\,673)
\footnote{Based on observations made with
a) the 60\,cm Bochum and the ESO 90\,cm Dutch telescopes, at La Silla, Chile, 
b) the 1.2\,m telescope at Kryonerion Observatory, Greece, 
c) the 1.3\,m telescope at Skinakas Observatory, Crete, Greece, and 
d) the European Space Agency OGS telescope operated on the island of
 Tenerife by the Instituto de Astrof\'{\i}sica
de Canarias in the Spanish Observatorio del Teide of the Instituto de 
Astrof\'{\i}sica de Canarias.}
}

\author{
D. Sinachopoulos\altaffilmark{2},
Th. Nakos\altaffilmark{3,4},
P. Boumis\altaffilmark{5,6},
E. van Dessel\altaffilmark{3},
M. Burger\altaffilmark{3},
\and
P. Rodr\'\i guez-Gil\altaffilmark{7}
}

\altaffiltext{2}{Institute of Astronomy \& Astrophysics, National Observatory of Athens,\\
      Lophos Kouphou\,-\,Palaia Penteli, GR - 15236, Greece}
\altaffiltext{3}{Royal Observatory of Belgium, Ringlaan 3, B-1180 Brussels, Belgium}
\altaffiltext{4}{Universit\'e de Li\`ege, Institut d'Astrophysique et de G\'eophysique, 
      Avenue de Cointe 5, B-4000 Li\`ege, Belgium}
\altaffiltext{5}{Department of Physics, University of Crete, PO Box 2208, 
      GR-710 03 Heraklion, Crete, Greece}
\altaffiltext{6}{Foundation for Research and Technology-Hellas, PO Box 1527, 
GR-711 10 Heraklion, Crete, Greece}
\altaffiltext{7}{Instituto de Astrof\'{\i}sica de Canarias, E-38200, La Laguna, Tenerife, Spain}

\begin{abstract}
We present the results of a photometric CCD monitoring of the gravitational
lens system UM\,673, that took place from 1995 to 2000. In total, the doubly-imaged quasar
was observed in the $R$-band during 29 photometric nights, using optical telescopes 
with dimensions in the range 0.6\,m to 1.3\,m. 
We detected a significant variability in the total light of the 
UM\,673 system, that is, in the light of the two QSO images plus the lensing 
galaxy. With respect to the magnitude of the gravitational lens system 
at its discovery, in 1986, 
UM\,673 was 0.3 magnitude brighter. Furthermore, our December 1996 measurements
show that between November 1995 and October 1997 the system became even brighter, reaching
a magnitude difference of 0.5 magnitude with respect to its discovery value. We also
present $R$~magnitudes and $V-R$~colours of seven field stars situated in the vicinity of 
the lens, based on a 3.5-month monitoring during the year 2000. 
\end{abstract}

\keywords{photometry, gravitational lensing, quasars: individual (0142-100)}

\section{Introduction}

\citet{surdej87,surdej88} presented UM\,673 ($0142-100$)
as a ``new case of gravitational lensing''. The two lensed images 
originate from a quasar at a redshift $z_{qso}\,=\,2.72$, while the 
lensing galaxy is much closer to the Milky Way ($z_l\,=\,0.49$).  
The angular separation between component $A$ and $B$ was found to be 
$\Delta\theta\,=\,2\farcs2$, while the magnitude of the QSO images
and the lens galaxy, at the time of discovery, was $R_A\,=\,16\fm9$,
$R_B\,=\,19\fm1$ and $R_g \approx 19.1$ magnitude, resulting
in a total magnitude $R_{A+B+G}\,=\,16.65$, with a standard deviation of 
the order of 0.2 magnitude. They concluded that UM\,673 appeared to be 
one of the most luminous quasars, due to the amplification
of image $A$ by gravitational lensing. 

A spectroscopic study of this gravitational lens was published by \citet{smette92},
in order to investigate the size of the corresponding $Ly\alpha$ clouds. 
Since they found that the equivalent widths of the $Ly\alpha$ lines in the spectra 
of $A$ and $B$ are well correlated,
they concluded that both light beams actually cross the 
same clouds. Data from this paper were used by \citet{miralda93}
to test their model for the minihalo of the $Ly\alpha$ forest
for the case of UM\,673.

\citet{daulie93} observed UM\,673 from 1987 to 1993. The observations were 
performed in the $B$-band, to eliminate any possible contamination originating
from the deflecting galaxy. They mainly derived the magnitude difference of the 
two QSO components, because the observing conditions were of modest quality for
most of the nights. The photometric zero point was computed for the nights of 
1988 October 31, 1988 December 17 and 1991 October 27. The respective
joint $B$ magnitude of the two QSO components (no contribution from the galaxy in the
$B$-band) was 16\fm8, 16\fm9 and 16\fm7. Their results 
led to the conclusion that the quasar was not showing any significant variability, 
but the accuracy of the relative photometry
was of the order of 0.05 to 0.2 magnitude. The mean $B$-band magnitude difference was  
$\Delta B\,=\,2\fm16\pm0\fm08$. In a later report based on an ESO Key-program
for photometric monitoring of gravitational lenses, \citet{courbin95} confirmed the above
conclusions.

During a 2-days workshop held in Hambourg, \citet{borgeest93} presented 
some {\sl preliminary} results from the 1988-1994 monitoring of the lens. 
The relative photometry of component $A$, with respect to a reference magnitude $R_o$, 
was indicating variations of the order of 0.2 magnitude! Unfortunately,
nothing more was ever published, and therefore no additional information concerning
the reduction method used or the reference magnitude could be found.    

In a recently published paper concerning $HST$ observations of gravitational
lenses, \citet{lehar00} reported on observations of UM\,673 performed in 
November 1994. The $R$ magnitude of the components $A$,~$B$, and the galaxy was
$16.67\pm0.06,18.96\pm0.05$ and $19.35\pm0.04$ respectively, 
resulting in an integrated magnitude of $R_{A+B+G}\,=\,16.47$. The error on the
total magnitude is of the order of $\sigma_R \approx 0\fm07$.

The Sloan Digital Sky Survey ($SDSS$) also observed UM\,673.
Having their own photometric system under preparation \citep{fukugita96}, 
they presented their measurements in a preliminary red filter,
designated as $r^{\star}$ 
(see section~\ref{comparison with the literature}).
Observations in the $r^{\star}$ filter were performed in 
1995 November 24 and 1996 December 12. For the first period,
\citet{richards97} reported $r^{\star}=16.47\pm0.01$, while for
the second period \citet{newberg99} reported $r^{\star}=16.52\pm0.01$. 

We report in this paper about a photometric monitoring through the years 1995-2000, 
with conclusions to significant variations in the total light of 
the gravitational lens system. We also present $R$ magnitudes
and $V-R$ colours of seven field stars, situated in the neighborhood of UM\,673.
These results, based on data which were accumulated during a 
period of 3.5 months, consist in a preliminary tool
for directly checking the photometric variability of the quasar. 

\section{Observations}

\subsection{Sites and instrumentation}

During the six-year monitoring period, we performed  CCD photometry 
of UM\,673 from four different observing sites. In detail, we observed at:

(a) La Silla (Chile) in November 1995, December 1996 and October 1997. During the 
first two periods, we used the 0.6 m (f/15) Bochum telescope, in conjunction with the 
TH7882 CCD (384$\times$576, 22 $\mu$m$^{2}$~pixels); in 1997,  
the observations were performed with the 0.9 m Dutch (f/11) telescope
and the Tektronix TK512CB CCD (512$\times$512, 27 $\mu$m$^{2}$~pixels). The above 
configurations gave a scale of 0.55$\arcsec$ pixel$^{-1}$~ and 
0.463$\arcsec$ pixel$^{-1}$~, corresponding to a field of 
view of 3.2$\arcmin \times 4.8\arcmin$ 
and $\sim 4\arcmin \times 4\arcmin$, respectively. All observations were carried out
through the standard Johnson $R$ filter. The exposure time for the first, 
second, and third period was 180, 300 and 1200 sec, respectively. 

(b) Kryonerion (Greece) in 1998 October 29, using the 1.2 m (f/13) 
telescope and the Class I SI502 CCD chip (512$\times$512, 24 $\mu$m$^{2}$~pixels), 
resulting in a scale of 0.30$\arcsec$ pixel$^{-1}$~with a field of view 
of 2.65$\arcmin \times 2.65\arcmin$. The images were taken using the $R$ filter,
with an integration time of 120 sec per exposure. A dark current 
subtraction was also included in the preliminary data reduction pipeline.

(c) Tenerife (Spain) in 2000 August 23, using the 1.0 m  OGS telescope 
(f/13.3 at the Ritchey-Chr\'etien) combined with a Thomson CCD 
(1024$\times$1024, 19$\mu$m$^{2}$~pixels) giving 
a scale of 0.3$\arcsec$ pixel$^{-1}$~and a field of view of 
5.12$\arcmin \times 5.12\arcmin$. This time the object was observed in both
the $V$ and $R$ filters, with a same exposure time of 120 sec.  

(d) Skinakas, Crete (Greece) from 2000 September to 2000 December, 
using the 1.3 m (f/7.7) Ritchey-Chr\'etien telescope. The SI003B CCD 
chip (1024$\times$1024, 24 $\mu$m$^{2}$~pixels) was the detector, 
giving a scale of 0.5$\arcsec$ pixel$^{-1}$ and 
a field of view of 8.57$\arcmin \times 8.57\arcmin$. The observations
were performed in the standard Johnson $V$ and $R$ filters,
with an exposure time of 240 sec per image.

Most of the nights UM\,673 was monitored for many hours and we have
obtained, in total, images from 29 clear nights. Individually, the median seeing 
at the Bochum telescope was 1\farcs5 and better at 
the Dutch telescope, around 1\farcs1. For the night 1998 October 29 
the seeing varied between 2\farcs0 - 3\farcs0, while on 2000 August 23
it was of sub-arsecond quality. Finally, a value between 0\farcs8 
to 1\farcs5 was measured for the period September\,-\,December 2000. 
The observational details are summarized in Table~\ref{logbook}.

\subsection{Photometric standards}
No study has ever been performed on neighbouring field stars to check the
variability of the quasar, since UM\,673 lies in a quasi-empty region of the sky. 
Therefore, the only possibility for almost 
all sets of observations (the Skinakas data were the single exception), was to use 
photometric standard stars (equatorial in our case) to calibrate our measurements 
(extinction and zero point correction). 

We mainly used the blue star HD\,12021 for the 1995\,-\,1998 missions. This 
photometric standard 
(HIPPARCOS identifier HIP 9155) has been extensively observed by different
observers \citep{klemola62, landolt83, cousins84, menzies91, grenon92} 
and its properties are known with high precision. Its equatorial co-ordinates
are: $\alpha_{2000}=01^h57^m56^s$ and $\delta_{2000}=-02\degr 06\arcmin 15\arcsec$;
\citet{menzies91} report: $V=8\fm877$ $\pm$0.015 and $(V-R_c)=-0\fm044$. 
In 2000 August 23 the calibration was performed by taking exposures 
at different airmasses of the
photometric standard 113191 \citep{landolt92}. For the Skinakas 
observations, apart from HD\,12021 we also used the Geneva photometric standard
with identification GEN +7.0093407 \citep{landolt92}.
>From the photometric standards observed in each mission a mean extinction coefficient
and zero point were computed for every (photometric) night, which were then
applied to the data of UM\,673. 

\section{Data reduction and results}

\subsection{Photometry of UM\,673}

The data reduction was performed by developing scripts in the ESO-MIDAS environment. 
We applied aperture photometry on our CCD frames in order to study
the photometric behavior of the {\sl total} light of UM\,673,
which is composed of the two lensed images and the lensing galaxy.
We used a circular aperture of 15\arcsec ~to 18\farcs5 diameter for this purpose, depending
on the observing run conditions (i.e. exposure time, quality of optics, seeing). 
The ``no-man's-land" was properly set, to avoid contamination from
the faint galaxy seen at the south of UM\,673. A precise 
calculation of the sky background near the gravitational lens
was performed by using a 2\farcs5 width ring, external to ``no-man's-land".
 
The incorporation of errors, apart from the aperture photometry errors,
included the uncertainty due to the extinction correction and the zero point
correction, as computed with the help of the photometric standard stars. 
The reduction of the data showed that the contribution of the first
factor on the daily averaged values was practically negligible 
(of the order of a few milli-magnitude), thanks to the numerous exposures and
the good observing conditions. The error due to the zero point and extinction
correction was of the order of 0\fm055, 0\fm04, 0\fm015, 0\fm04 
and 0\fm015 for the 1995 to 2000 missions, respectively. 

Table~\ref{photometricresults} contains the $R$-band photometric results. 
Column 2 presents the daily averages for each run, with their accuracy, 
composed of all exposures of each night. Column 3 contains the mean value and its
standard deviation for each run. Our results, together with
the two $R$-band measurements available in the literature 
(\citet{surdej88}, \citet{lehar00}) are plotted in Fig.~\ref{variability}. 
The variability of the gravitational lens system is clearly seen in the graph. 

\subsection{Reference stars in the vicinity of UM\,673}

Thanks to the relatively large field of view of the configuration of the Skinakas 
1.3\,m telescope, it was possible to perform aperture photometry on ten field
stars in the neighborhood of UM\,673, which were present in the majority of our frames. 
Although almost all of them are fainter 
than the lens, an optimized combination of
several factors (i.e. photometric conditions, high quality telescope optics,  
precise tracking by means of an autoguider) resulted in a large set of 
good quality data (81 frames in $R$ and 60 frames in $V$, obtained during a period of
3.5 months), from which significant information on these field stars
could be extracted. 

The variability of the field stars was checked by properly co-adding all 
the frames of the given night and then performing aperture photometry on them. 
This operation gave a set of eight frames per filter. For every
frame we computed the magnitude difference, $\Delta m_{ik}$, of the
star $k$ with respect to the star $i$ and finally obtained a mean magnitude 
difference, $\Delta \bar{m}_{ik}$, for every pair of stars in the whole data set. 
The results obtained from the differential photometry showed that 
for most cases the mean errors in the magnitude differences were 
between 0.01-0.03 magnitude, except for three stars.
Therefore, we finally decided to reject them, and to use the rest
for checking the variability of the lens.

\begin{figure}
\plotone{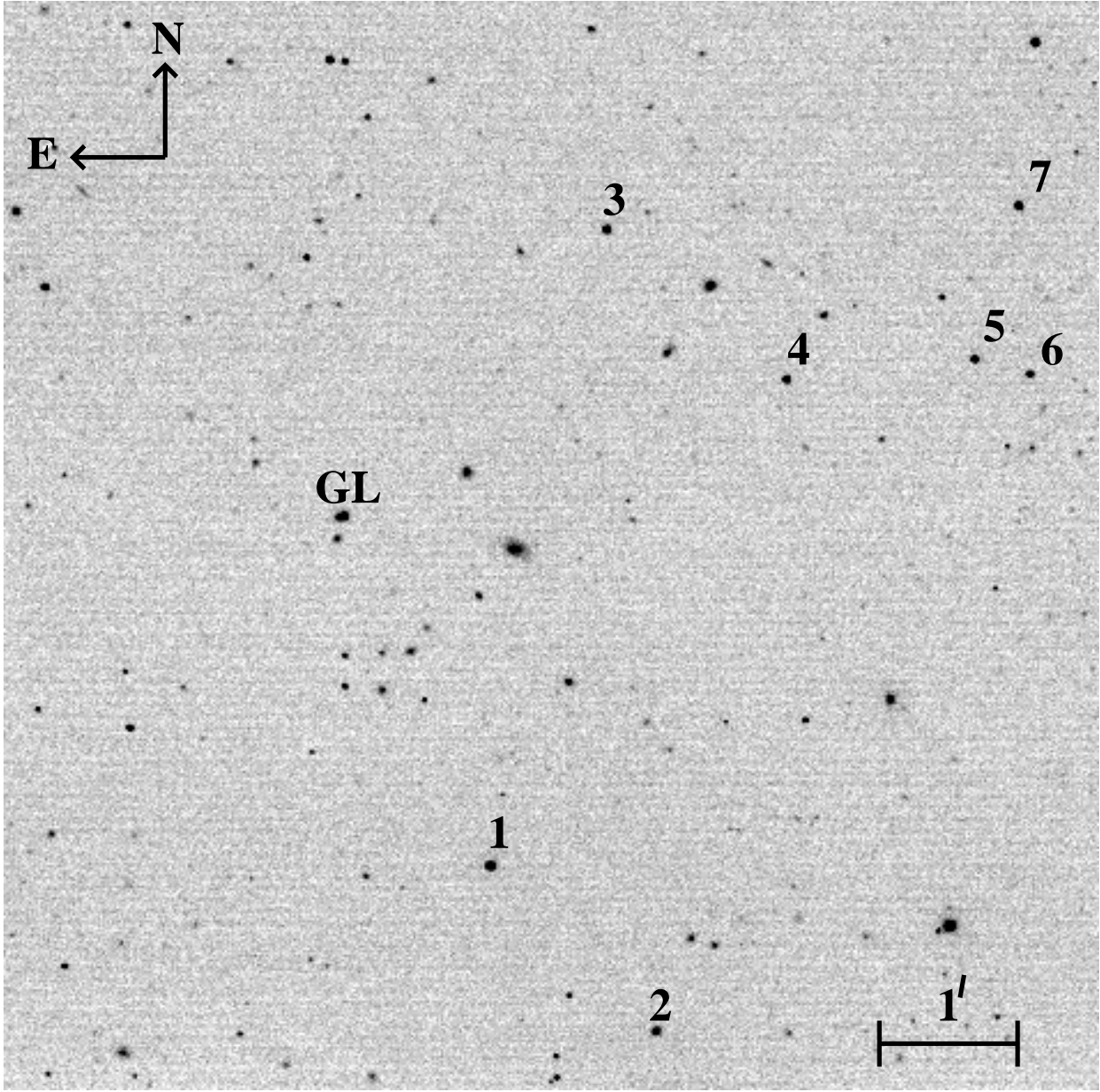}
\caption{Finding chart (8$\arcmin \times 8\arcmin$) of UM\,673 in the $R$-band, 
where the reference field stars (labeled 1$-$7) are shown. The position of 
the lens system is also marked. This figure consists of four different 
combined images resulting in a total exposure time of 16 min 
(4 min each). \label{findingchart} }
\end{figure}

\setcounter{footnote}{0}
A finding chart with the identification of the field 
stars is given in Fig.~\ref{findingchart}. By using the same photometric standards
that were used for the lens, we performed a zero-point and extinction
correction, converting the instrumental values to $V$ and $R$ magnitudes. 
Second order extinction correction and colour transformation were not performed.
We also performed a preliminary 
astrometric identification with the help of the $XEphem$ software, developed by
E. Downey (2000)\footnote{{\it $XEphem$\,-\,An Interactive Astronomical Ephemeris Program for
$X$ Windows} is available at http://www.ClearSkyInstitute.com/xephem}. By correlating 
our results with the co-ordinates presented in \citet{newberg99}, we  
identified the seven field stars.  Table~\ref{referencestars} 
lists the RA-Dec co-ordinates of the reference stars, their $R$~magnitude and $(V-R)$~colour, 
including errors associated with the determination of the instrumental magnitudes. 

\section{Discussion on the variability of the lens}

\label{comparison with the literature}
Although it is approximately fifteen years since the discovery of the lens,
UM\,673 has hardly been observed. Consequently, the very few existing 
measurements can not provide a clear view concerning the
photometric behavior of the gravitational lens system before 1995. 

The closest (in time) observations to our measurements are those performed by
the $SDSS$, with a photometric system consistsing of five passbands ($u',g',r',
i'$ and $z'$). Since the definition of the system is 
still under progress, the observations are performed using 
filters which are similar but {\it not} identical to the final ones,
designated as $u^{\star},g^{\star},r^{\star},i^{\star}$ and $z^{\star}$. 

UM\,673 was observed by the $SDSS$ group in 1995 November 24, 
and in 1996 December 12. For the first period, \citet{richards97}
report $u^{\star}=17.35\pm0.07, g^{\star}=16.63\pm0.01, 
r^{\star}=16.47\pm0.01$, while for the second period 
\citet{newberg99} report $17.30\pm0.01, 16.70\pm0.01$ and $16.52\pm0.01$,
respectively. The transformation of the $SDSS$ measurements to the
Johnson $UBV$ and Cousins $RI_c$ photometric system is a complicated 
task (cf \citet{krisciunas98} and \citet{newberg99}) and, as
a consequence, a direct comparison of the $SDSS$ with our measurements 
is not possible. Nevertheless, since both we and the $SDSS$ have data 
obtained during the same periods (November 1995 and December 1996), 
we tried to derive some additional conclusions on the variability 
of the lens by comparing the observations obtained by $SDSS$  
in the two different periods.

The comparison of the 1995 and 1996 $SDSS$ measurements yielded
ambiguous results, since the decrease of the $u^{\star}$
magnitude shows that the lens became brighter, while a comparison
of the $g^{\star}$ and $r^{\star}$ measurements gave the
opposite conclusion. As it is mentioned in \citet{newberg99}, 
the main task of the 1996 $SDSS$ observations was to explore
the detection of QSOs, based on colour selection criteria. 
For this reason, observations were performed even if the conditions
were non-photometric. As a consequence, we conclude that  
a comparison of the 1995 and 1996 $SDSS$ measurements 
does not give any supplementary information on the variability 
of the quasar.

The instrumentation used during our 1995-1996 missions did
not allow us any additional exploitation of the
data (e.g. profile fitting of the QSO components). 
Hence, both a microlensing scenario and an 
internal variability of the quasar
could equivalently explain the brightening of UM\,673. 
However, our measurements, which are characterized by a
photometric precision of the order of a few 
hundredths of a magnitude, clearly demonstrate the 1996 
peak in brightness. At the same time, they indicate that since 
1997 the quasar seems to be in a less variable state, but 
the situation of a quiet phase can not be ruled out.

\section{Conclusions}

We have been monitoring the gravitational lens UM\,673 through the 
years 1995\,-\,2000. To measure the total light of the lens  
(component $A$ + component $B$ + lensing galaxy) we constrained
ourselves to aperture photometry. The numerous observations obtained 
during each run (several hours per night), allowed us to reach a 
photometric precision of a few hundredths of a magnitude. 
The photometric calibration was
performed with the help of several photometric standard stars.
Although the observations were performed using several 
configurations, all measurements agree with the conclusion 
that the QSO has undergone a long term variability,
 that corresponds to a brightening in the total light of the system of 
$\approx$ 0.3 magnitude in respect to its discovery value, in
1987. From the agreement between the 1995 and 1997 measurements we see that
this variability had its peak (additional 0\fm2 brightening) in 1996.
 
The $R$ magnitude of UM\,673 we derived was 16.30$\,\pm\,$0.02 in November 1995,  
16.08$\,\pm\,$0.01 in December 1996, 16.30$\,\pm\,$0.01 in October 1997, 
16.23$\,\pm\,$0.05 in October 1998 and 16.34$\,\pm\,$0.01 for the last 
quarter of 2000 (Fig. \ref{variability}). 

Thanks to the optimized configuration of the Skinakas observatory, we managed
to have $V$ and $R$ measurements of seven field stars in the 
vicinity of UM\,673 during the period September\,-\,December 2000. 
Although more observations are needed for a better knowledge 
about the stability of these stars over a longer time-basis,
the photometric reduction showed no significant variability at the
level 0\fm02\,-\,0\fm03.
 
The measurements obtained after 1996 seem to form a plateau at
around $R\approx16.3$, indicating that since 1997 the quasar seems
to be in a less variable, or even quiet, state. Nevertheless, 
we propose a systematic monitoring of UM\,673 at 
a rate of a few images at fixed time intervals. With the help
of the reference stars, differential photometry would directly show
any new variability in the system. We also suggest new measurements
with instruments allowing analysis of the spectra of the components 
for comparison with earlier ones, and photometric observations to
derive the fluxes of the two QSO components.

\acknowledgments We thank Dr. A. Erikson (DLR, Berlin) for his 
introduction to the 0.6\,m Bochum 
telescope and for his help to convert our CCD data to FITS format, and 
A. Kougentakis for his excellent assistance during the observations at 
Skinakas Observatory. T.N. would like to thank N. Shatsky, 
visiting astronomer at ROB, for the helpful discussions. \\
The authors would like to acknowledge Prof. J. Surdej, for his 
useful comments. Finally, it is a pleasure to thank the referee, 
Dr. R. Schild, for his constructive suggestions.\\
This research (D.S.) was mainly carried out in the framework of the projects
``Service Centres and Research Networks'', and 
``P\^oles d'Attraction Interuniversitaires'' P4/05,
both initiated and financed by the Belgian Federal Scientific Services (DWTC/SSTC). 
T.N. acknowledges support from the project  "Chercheurs Suppl\'ementaires aux 
Etablissements Scientifiques Federaux".
P.B. acknowledges support from a ``P.EN.E.D." program of 
the General Secretariat of Research and Technology of Greece, and the 
Royal Observatory of Belgium for financing his visit in Brussels. \\
Skinakas Observatory is a collaborative project of the University of 
Crete, the Foundation for Research and Technology-Hellas and the 
Max-Planck-Institut f\"{u}r Extraterrestrische Physik.

\clearpage

\begin{figure}
\plotone{Sinachopoulos.fig1.eps}
\caption[]{$R$-band variability of UM\,673 
during a fourteen-year interval (total light). The symbols ($\bullet$) ($\blacktriangle$) 
and ($\blacksquare$) correspond to \citet{surdej88}, \citet{lehar00}
and our measurements, respectively. 
\label{variability} }
\end{figure}

\clearpage

\clearpage

\begin{deluxetable}{rrrrr} 
\tablecolumns{5} 
\tablewidth{0pc} 
\tablecaption{Observational details \label{logbook}} 
\tablehead{  
\colhead{Date} & \colhead{Instrument} & \colhead{Detector} & \colhead{Filter} & \colhead{Exposure time}\\
\colhead{} & \colhead{(telescope)} & \colhead{(CCD type)} & \colhead{} & \colhead{(sec)}}
\startdata
1995 Nov 03/09-10 & 0.6\,m Bochum\phn & TH 7882\phm{space.} & R\phm{te} & 180\phm{space}\\ 
1996 Dec 09-19\phm{tex.} & 0.6\,m Bochum\phn & TH 7882\phm{space.} &  R\phm{te} & 300\phm{space}\\
1997 Oct 22-26\phm{tex.} & 0.9\,m Dutch\phm{te} & Tektronix TK512CB &  R\phm{te} & 1200\phm{space}\\
1998 Oct 29\phm{spaces.} & 1.2\,m Kryonerion & Class I SI502\phm{tex} & R\phm{te} & 120\phm{space} \\
2000 Aug 23\phm{space1} & 1.0\,m OGS\phm{tex} & Thomson\phm{space} & R\phm{te} & 120\phm{space} \\
2000 Sep 03/11/18 & 1.3\,m Skinakas\phm{t} & Tektronix SI003B\phn & R\phm{te} & 240\phm{space} \\
2000 Nov 02-03/23 & 1.3\,m Skinakas\phm{t} & Tektronix SI003B\phn & R\phm{te} & 240\phm{space} \\
2000 Dec 14-15\phm{tex.} & 1.3\,m Skinakas\phm{t} & Tektronix SI003B\phn & R\phm{te} & 240\phm{space}\\
\enddata
\end{deluxetable}

\clearpage

\begin{deluxetable}{rrr} 
\tablecolumns{3} 
\tablewidth{0pc} 
\tablecaption{Mean daily total $R$ magnitudes of the UM\,673 system and 
its errors are presented in column 2. Column 3 provides the average value of each
run and its standard deviation.\label{photometricresults}} 
\tablehead{  
\colhead{Date} & \colhead{$R$} & \colhead{$\bar{R}$}}
\startdata
1995 Nov 03 & 16.27 $\pm$ 0.06 & \\
1995 Nov 09 & 16.31 $\pm$ 0.06 & 16.30 $\pm$ 0.02 \\
1995 Nov 10 & 16.32 $\pm$ 0.06 & \\ 
\tableline 
1996 Dec 09 & 16.11 $\pm$ 0.05 & \\
1996 Dec 10 & 16.12 $\pm$ 0.05 & \\
1996 Dec 11 & 16.10 $\pm$ 0.05 & \\
1996 Dec 12 & 16.09 $\pm$ 0.05 & \\
1996 Dec 13 & 16.05 $\pm$ 0.05 & \\
1996 Dec 14 & 16.07 $\pm$ 0.05 & 16.08 $\pm$ 0.01 \\
1996 Dec 15 & 16.08 $\pm$ 0.05 & \\
1996 Dec 16 & 16.11 $\pm$ 0.05 & \\
1996 Dec 17 & 16.07 $\pm$ 0.05 & \\
1996 Dec 18 & 16.08 $\pm$ 0.05 & \\
1996 Dec 19 & 16.04 $\pm$ 0.05 & \\
\tableline
1997 Oct 22 & 16.29 $\pm$ 0.02 & \\
1997 Oct 23 & 16.30 $\pm$ 0.02 & \\
1997 Oct 24 & 16.30 $\pm$ 0.02 & 16.30 $\pm$ 0.01 \\
1997 Oct 25 & 16.30 $\pm$ 0.02 & \\
1997 Oct 26 & 16.29 $\pm$ 0.02 & \\ 
\tableline 
1998 Oct 29 & 16.23 $\pm$ 0.05 & 16.23 $\pm$ 0.05 \\ 
\tableline
2000 Aug 23 & 16.30 $\pm$ 0.02 & \\
2000 Sep 03 & 16.34 $\pm$ 0.02 & \\
2000 Sep 11 & 16.35 $\pm$ 0.02 & \\
2000 Sep 18 & 16.31 $\pm$ 0.03 & \\
2000 Nov 02 & 16.35 $\pm$ 0.02 & 16.34 $\pm$ 0.01 \\
2000 Nov 03 & 16.34 $\pm$ 0.02 & \\
2000 Nov 23 & 16.35 $\pm$ 0.02 & \\
2000 Dec 14 & 16.34 $\pm$ 0.02 & \\
2000 Dec 15 & 16.35 $\pm$ 0.02 & \\
\enddata 
\end{deluxetable}

\clearpage

\begin{deluxetable}{rrrr} 
\tablecolumns{4} 
\tablewidth{0pc} 
\tablecaption{New reference stars in the field of UM\,673.\label{referencestars}} 
\tablehead{  
\colhead{N$^{o}$ of star} & \colhead{RA (2000)  Dec (2000)} & \colhead{$R$} & \colhead{$V-R$} \\
\colhead{} & \colhead{(h\phn m\phn s)\phm{text}($\degr~~\arcmin~~\arcsec$)} & \colhead{(mag)} & \colhead{(mag)}}
\startdata
1\phm{text} & 01 45 12.05\phm{te} $-$\,09 47 49\phm{.} & 15.92 $\pm$ 0.02 & 0.51 $\pm$ 0.03 \\
2\phm{text} & 01 45 07.15\phm{te} $-$\,09 49 01\phm{.} & 16.61 $\pm$ 0.02 & 0.94 $\pm$ 0.03 \\
3\phm{text} & 01 45 08.73\phm{te} $-$\,09 43 13\phm{.} & 17.12 $\pm$ 0.02 & 0.33 $\pm$ 0.03 \\
4\phm{text} & 01 45 03.43\phm{te} $-$\,09 44 18\phm{.} & 17.37 $\pm$ 0.02 & 0.84 $\pm$ 0.04 \\
5\phm{text} & 01 44 57.90\phm{te} $-$\,09 44 09\phm{.} & 17.42 $\pm$ 0.02 & 0.84 $\pm$ 0.04 \\
6\phm{text} & 01 44 56.26\phm{te} $-$\,09 44 15\phm{.} & 17.80 $\pm$ 0.02 & 0.42 $\pm$ 0.04 \\
7\phm{text} & 01 44 56.63\phm{te} $-$\,09 43 02\phm{.} & 16.75 $\pm$ 0.03 & 0.38 $\pm$ 0.04 \\
\enddata
\end{deluxetable}

\end{document}